\documentclass[a4paper,11pt]{article}
\pdfoutput=1 

\usepackage{jheppub} 
\usepackage{graphicx}  
\usepackage{dcolumn}   
\usepackage{bm}        
\usepackage{amssymb}   
\usepackage{braket}
\usepackage{amsmath}
\usepackage{color}  	

\usepackage{bbm}
\usepackage{slashed}
\usepackage{subfigure}
\usepackage[utf8]{inputenc}
\DeclareGraphicsExtensions{.pdf,.png,.jpg,.eps}
\graphicspath{ {./images/} }

\newcommand{\Tr}{\mathrm{Tr}}
\newcommand{\Pexp}{\mathrm{Pexp}}
\newcommand{\td}{\mathrm{d}}

\newcommand{\TT}[1]{\mathrm{#1}}

\renewcommand{\v}[1]{\mathbf{#1}}

\renewcommand{\>}{\right\rangle}

\newcommand{\lkl}{\left|}

\newcommand{\qqquad}{\qquad\qquad}
\newcommand{\qqqquad}{\qquad\qquad\qquad}
\newcommand{\qqqqquad}{\qquad\qquad\qquad\qquad}

\newcommand{\As}{\alpha_{\mathrm{s}}}

\begin{document}

\title{Coulomb gluons will generally destroy coherence}
\author[a]{Jeffrey R. Forshaw,}
\author[a,b]{Jack Holguin}

\affiliation[a]{Consortium for Fundamental Physics, School of Physics \& Astronomy, \\
	University of Manchester, Manchester M13 9PL, United Kingdom}
\affiliation[b]{CPHT, CNRS, Ecole polytechnique, IP Paris, F-91128 Palaiseau, France}

\emailAdd{jeffrey.forshaw@manchester.ac.uk}
\emailAdd{jack.holguin@polytechnique.edu}

\date{\today}

\abstract{Coherence violation is an interesting and counter-intuitive phenomenon in QCD. We discuss the circumstances under which violation occurs in observables sensitive to soft radiation and arrive at the conclusion that almost all such observables at hadron colliders will violate coherence to some degree. We illustrate our discussion by considering the gaps-between-jets observable, where coherence violation is super-leading, then we generalise to other observables, including precise statements on the logarithmic order of coherence violation.}

\maketitle

\section{Introduction}
The collinear evolution of hadronic parton densities is accounted for using the equations of Dokshitzer, Gribov, Lipatov, Altarelli \& Parisi (DGLAP) \cite{DOKSHITZER1980269,Gribov:1972ri,APSplitting}. Inspired by the classic factorisation proofs in \cite{Collins:1987pm,Collins:1988ig}, it is tempting to assume that, as a result of QCD coherence, this collinear evolution can always be factorised from any wide-angle, soft-gluon emissions. However, Coulomb/Glauber exchanges can destroy coherence and invalidate the factorisation \cite{Catani:2011st,factorisationBreaking} with potentially important phenomenological consequences \cite{Becher:2021zkk}. In this paper, we explore the circumstances under which this happens.

\subsection{Case study: gaps between jets}
Oderda \& Sterman (OS) \cite{Kidonakis:1998nf,Oderda:1998en,Oderda:1999kr} presented the first calculations of the rate for the production of two or more jets subject to the restriction that there should be no additional jets located in the rapidity interval between the two highest $p_T$ jets (the dijets) with transverse momentum (or energy) bigger than some value, $Q_0$. This observable is sensitive to logarithmically enhanced, wide-angle, soft-gluon emissions. According to OS, the leading logarithmic (LL) contribution to the gaps-between-jets cross-section at a hadron collider is
\begin{equation}
	\frac{\td \sigma_\mathrm{OS}}{\td x_a \td x_b\, \td \mathcal{B}} = f_A(x_a,Q) f_B(x_b,Q) \text{Tr} ( \v{V}_{Q_0,Q} \v{H} \v{V}_{Q_0,Q}^\dag), \label{eq:OS1}
\end{equation} 
where $\v{H} = |M_0 \rangle \langle M_0|$ is the QCD hard scattering matrix ($|M_0\rangle$ is the lowest-order, QCD hard-process amplitude for dijet production), $\td \mathcal{B} \equiv \td y \, \td^{2} \mathbf{p}_{\bot} / 16 \pi^2 \hat{s}$ is the measure for the on-shell (Born) kinematics of the final state dijets, $f_{A/B}$ are parton distribution functions for the incoming hadrons $A$ and $B$, $Q$ is the jet transverse momentum and 
\begin{equation}
\v{V}_{Q_0,Q} \approx \exp \left( -\frac{\alpha_s}{\pi} \ln \frac{Q}{Q_0} (Y \v{T}_t^2 + i \pi \v{T}_s^2) \right). \label{eq:sudakov}
\end{equation}
The rapidity separation between the dijets is $Y$, $\v{T}_t^2$ is the colour-space operator corresponding to the colour exchanged in the $t$-channel and $\v{T}_s^2$ corresponds to the colour exchanged in the $s$-channel. For example, if the hard process is $ab \to cd$ then $\v{T}_s^2 = (\v{T}_a+\v{T}_b)^2 = (\v{T}_c+\v{T}_d)^2$ and $\v{T}_t^2 = (\v{T}_a+\v{T}_c)^2=(\v{T}_b+\v{T}_d)^2$. The Sudakov operator $\v{V}_{Q_0,Q}$ corresponds to no soft-gluon emission directly into the region between the dijets with transverse momentum greater than $Q_0$. Eq.~\eqref{eq:sudakov} is a good approximation for $Y \gg 1$ (the terms we have neglected are proportional to the unit matrix in colour space, which means their neglect does not affect what follows). 

Following the discovery of non-global logarithms by Dasgupta \& Salam \cite{non_global_logs}, it became clear that the OS analysis was incomplete because it did not account for the Sudakov suppression associated with partons originally radiated into the out-of-gap region. Including this physics makes the problem considerably more complicated/interesting. Notwithstanding the role of this non-global radiation, our focus here is on the collinear evolution of the incoming partons and for that we will continue to neglect the non-global corrections. We do so for pedagogical reasons, fully aware that non-global corrections are important. That said, let us return to Eq.~\eqref{eq:OS1} and notice that it is still not quite right. 

To see what is wrong, let us consider only the order $\alpha_s$ correction to the collinear evolution of the parton densities above the veto scale $Q_0$ \cite{Forshaw:2006fk,Forshaw:2008cq,Forshaw:2019ver}. For simplicity, we only consider quark evolution from hadron $A$. The result is\footnote{This is the first term in the `out-of-gap' expansion introduced in \cite{Forshaw:2006fk,Forshaw:2008cq}.}
\begin{align}
	\frac{\td \sigma_1}{\td x_a \td x_b\, \td \mathcal{B}} =& \frac{\alpha_s}{\pi} \int_{Q_0}^Q \frac{\td k_T}{k_T} \int_{0}^{1-\frac{k_T}{Q}e^{Y/2}} \frac{\td z}{z} \, P_{qq}(z) \,  f_B(x_b,Q) \nonumber \\ & \times \Bigg[ \Theta (z - x_a) \, f_A(x_a/z,Q_0) \frac{1}{\v{T}_a^2} \text{Tr} (\v{V}_{Q_0,k_T} \v{T}_a\v{V}_{k_T,Q} \v{H}\v{V}_{k_T, Q}^\dag \v{T}_a^\dag\v{V}_{Q_0,k_T}^\dag) \nonumber \\ & \qqqqquad \qqqquad \quad - z f_A(x_a,Q_0) \text{Tr} (\v{V}_{Q_0,Q} \v{H} \v{V}_{Q_0, Q}^\dag)  \Bigg] \label{eq:master}~.
\end{align}
Here $\v{T}_a^2 = C_F$ since we are supposing that parton $a$ is a quark and we are following convention in writing
\begin{equation}
	P_{qq} = C_F \frac{1+z^2}{1-z}.
\end{equation} 
In the case of no jet veto we have the familiar result for the dijet cross-section including the evolution of the parton distribution functions in terms of the plus prescription \cite{DOKSHITZER1980269,Gribov:1972ri,APSplitting,Dokshitzer:1991wu}:
\begin{align}
	\frac{\td \sigma_1}{\td x_a \td x_b\, \td \mathcal{B}} &= \frac{\alpha_s}{\pi} \int_{Q_0}^Q \frac{\td k_T}{k_T} \int_{0}^{1-\frac{k_T}{Q}} \frac{\td z}{z} \, P_{qq}(z)  \Big[  \Theta (z - x_a) \, f_A(x_a/z,Q_0) \nonumber \\ & \qqqqquad \qqqquad - z f_A(x_a,Q_0) \Big] \text{Tr} (\v{H}) f_B(x_b,Q), \nonumber \\
	&= \frac{\alpha_s}{\pi} \int_{Q_0}^Q \frac{\td k_T}{k_T} \int_{x_a}^{1} \frac{\td z}{z} \, C_F \left(\frac{1+z^2}{1-z}\right)_+  \, f_A(x_a/z,Q_0) f_B(x_b,Q)  \text{Tr} (\v{H}) ~. \label{eq:DDT}
\end{align}
In the first line, it is safe to set the upper limit of the $z$ integral to unity since the integrand vanishes in that limit. This allows us to write the second equality in terms of the plus prescription.

The problem in Eq.~\eqref{eq:master} is the non-commutativity of the colour emission operator $\v{T}_a$ with the Sudakov operator, which can be traced to the Coulomb/Glauber $i \pi$ term in Eq.~\eqref{eq:sudakov}. In the case of the gaps-between-jets observable, expanding Eq.~\eqref{eq:master} order-by-order in $\alpha_s$ reveals an unexpected double logarithmic enhancement starting at $\sim \pi^2 N_c^2 Y \alpha_s^4 \log^5(Q/Q_0)$ relative to the inclusive dijet cross section. These are the super-leading logarithms first reported in \cite{Forshaw:2006fk}. Notice that the $i \pi$ terms do not cancel or factorise because $[\v{T}_t^2, \v{T}_s^2] \neq 0$ and $[\v{T}_a, \v{T}_s^2] \neq 0$. This is what spoils our ability to factorise the collinear evolution into the parton distribution functions since, in the absence of any $i \pi$ terms, we would recover Eq.~\eqref{eq:OS1} due to the fact that $[\v{T}_a, \v{T}_t^2] = 0$. The problem concerns only emissions collinear to one of the two incoming partons since emissions collinear to the outgoing partons occur long after the hard scattering and cannot therefore influence the colour dynamics. As a result of this causal structure, $[\v{T}_{c,d}, \v{T}_s^2] = 0$ and $[\v{T}_{c,d}, \v{T}_t^2] = 0$ and final-state collinear emissions do factorise.

\section{General considerations}

We now consider more general pure QCD processes in hadron-hadron collisions (we will consider electroweak processes later) and begin with a generalisation of Eq.~\eqref{eq:master}:
	\begin{align}
	\frac{\td \sigma_1}{\td x_a \td x_b \, \td \mathcal{B}} =& \frac{\alpha_s}{\pi} \int_{\mu_{\TT{F}}}^Q \frac{\td k_{\bot}}{k_{\bot}} \int_{0}^{1-k_{\bot}/Q} \frac{\td z}{z} \, P_{qq}(z) \; u_{1}(k) \, f_B(x_b,Q) \nonumber \\ 
	& \times \Bigg[ \Theta (z - x_a) \, f_A(x_a/z,\mu_{\TT{F}}) \frac{1}{\v{T}_a^2} \text{Tr} (\v{V}_{\mu_{\TT{F}},k_{\bot}} \v{T}_a\v{V}_{k_{\bot},Q} \v{H}\v{V}_{k_{\bot}, Q}^\dag \v{T}_a^\dag\v{V}_{\mu_{\TT{F}},k_{\bot}}^\dag) \nonumber \\ & \qqqqquad \qqqquad - z f_A(x_a, \mu_{\TT{F}}) \text{Tr} (\v{V}_{\mu_{\TT{F}},Q} \v{H} \v{V}_{\mu_{\TT{F}}, Q}^\dag)  \Bigg] , \label{eq:start}
	\end{align}
where $Q$ is the hard scale and 
	\begin{align}
	\v{V}_{\alpha,\beta} \approx \Pexp \Bigg( \frac{\As}{\pi}   \Bigg[ \sum_{i \neq j} \v{T}_{i} \cdot \v{T}_{j} \int_{\alpha}^{\beta} \frac{\td  q^{(ij)}_{\bot}}{q^{(ij)}_{\bot}}  \int_{-\ln Q/q^{(ij)}_\bot}^{\ln Q/q^{(ij)}_\bot} \td y^{(ij)} \int_0^{2\pi}\frac{\td \phi^{(ij)}}{4\pi} &\big(1 - u_{n}(q,\{k\}_{n-1})\big) \nonumber \\ & - i\pi \v{T}_s^2 \ln \frac{b}{a} \Bigg] \Bigg). \label{eq:V}
	\end{align}
Here the limits on $z$ and $y^{(ij)}$ are purely kinematic. The function $u_n$ is related to the measurement function $u$, such that
\begin{align}
    \sigma = \sum_{n} \int \td \sigma_{n} \; u(k_{1},\dots, k_{n}),
\end{align}
where $\sigma$ is the observable cross-section, $n$ is the number of emissions with respect to the Born process and momenta are written with the largest $k_{\bot}$ to the left. We define $u_{n}$ so that $$u(k_{1},\dots, k_{n}) \equiv u_{n}(k_{n},\{k\}_{n-1}) u(k_{1},\dots, k_{n-1}),$$ where $\{k\}_{n} =  \{k_{1},\dots, k_{n}\}$. In Eq.~\eqref{eq:start} we only need $u(q)  \equiv u_1(q)$ and $u(k,q) \equiv u_2(q,k) u_1(k)$. In Eq.~\eqref{eq:V}, 
$$\big( q^{(ij)}_{\bot} \big)^{2} = \frac{2 k_{i}\cdot q \, k_{j}\cdot q}{k_{i}\cdot k_{j}}$$
is the transverse momentum defined in the zero momentum frame of partons $i$ and $j$; $y^{(ij)}$ and $\phi^{(ij)}$ are the rapidity and azimuth in the same frame. The sum over $i$ and $j$ in Eq.~\eqref{eq:V} is over all prior real emissions and as such it is context dependent.

Eq.~\eqref{eq:start} will generate coherence violating terms at some perturbative order if the Coulomb terms do not entirely cancel. For this cancellation to occur we require
\begin{align}
[ \TT{Re} (\ln \v{V}_{\mu_{\TT{F}},k_{\bot}}) , \v{T}^{2}_{s}] = 0. \label{eq:noCV}
\end{align}
This is because if Eq.~\eqref{eq:noCV} is satisfied we can write $\v{V}_{\mu_{\TT{F}},k_{\bot}} = \v{V}^{\TT{Re}}_{\mu_{\TT{F}},k_{\bot}} \v{V}^{\TT{Im}}_{\mu_{\TT{F}},k_{\bot}}$ where $\v{V}^{\TT{Im}}_{\mu_{\TT{F}},k_{\bot}} =  e^{\TT{Im} ( \ln \v{V}_{\mu_{\TT{F}},k_{\bot}} )}$ and  $(\v{V}^{\TT{Im}}_{\mu_{\TT{F}},k_{\bot}})^{\dag} = (\v{V}^{\TT{Im}}_{\mu_{\TT{F}},k_{\bot}})^{-1}$. This permits the cancellation of the outermost Coulomb terms and then, since 
\begin{align}
[ \TT{Re} (\ln \v{V}_{\mu_{\TT{F}},k_{\bot}}) , \v{T}_{a}] = 0,
\end{align}
a cascade effect leads to the cancellation of all other Coulomb terms \cite{Collins:1987pm,Collins:1988ig,Forshaw:2019ver}. 

Eq.~\eqref{eq:noCV} can be generalized to a statement that there be no coherence violation in $\sigma_n$, i.e. for any number of collinear emissions, thereby allowing all-orders DGLAP evolution up to the hard scale $Q$. For this to be so, it is necessary that 
\begin{align}
[ \TT{Re} (\ln \v{V}_{\alpha,\beta}) , \v{T}^{2}_{s}] \lkl \mathcal{M}_{0}^{(n)} \>= 0 \label{eq:noCVallO},
\end{align}
where $\lkl \mathcal{M}_{0}^{(n)} \>$ is the Born amplitude dressed with $n$ soft or collinear partons.
Eq.~\eqref{eq:noCVallO} means that 
\begin{align}
\bigg[ \sum_{i \neq j} \v{T}_{i} \cdot \v{T}_{j} \Omega_{ij} \, , \v{T}^{2}_{s}\bigg] & = \bigg[ \bigg(\sum_{i = a,b}+\sum_{i \neq a,b}\bigg)\bigg(\sum_{j = a,b} + \sum_{j \neq a,b}\bigg) \v{T}_{i} \cdot \v{T}_{j} \Omega_{ij} \, , \v{T}^{2}_{s}\bigg] \nonumber \\
&= 2 \bigg[ \sum_{i = a,b}\sum_{j \neq a,b} \v{T}_{i} \cdot \v{T}_{j} \Omega_{ij}\, , \v{T}^{2}_{s}\bigg] = 0, \label{eq:noCVhh}
\end{align}
where $\TT{Re} (\ln \v{V}_{\alpha,\beta}) \propto \sum_{i \neq j} \v{T}_{i} \cdot \v{T}_{j} \Omega_{ij}$ and it is understood that the commutators are to act on $\lkl \mathcal{M}_{0}^{(n)} \>$. In other words, we only need to check the commutativity of a Coulomb exchange with any soft interference term between an initial and a final state parton in order to check for coherence violation. 

For processes with fewer than two coloured, incoming particles or when the incoming particles form a colour singlet Eq.~\eqref{eq:noCVhh} is automatically satisfied since $\v{T}_s^2$ is a Casimir. For all other processes, the commutator in Eq.~\eqref{eq:noCVhh} only vanishes if $\Omega_{aj} = \Omega_{bj}$\footnote{ The commutator also vanishes if $\Omega_{ij} = \Omega_{ij'}$ for all $j,j' \not\in \{a,b\}$ and $i\in \{a,b\}$. However this is kinematically impossible when $j$ is hard and $j'$ is soft.}. This is the case for double logarithmic terms in $\Omega_{aj}$ however it is not the case for single logarithmic, wide-angle, terms\footnote{This is because in the double logarithmic approximation $q^{(aj)}_{\bot}\approx q^{(bj)}_{\bot}$. }. Quite generally, 
\begin{align}
\Omega_{a j} =  \int \frac{\td q^{(ab)}_{\bot}}{ q^{(ab)}_{\bot}} \int   \td y^{(ab)}\td \phi^{(ab)} & \big(q^{(ab)}_{\bot}\big)^2 \frac{k_{a}\cdot k_{j}}{ k_{a}\cdot q \, k_{j}\cdot q} \times (1-u_{n}(q,\{ k \}_{n-1})) \nonumber \\ & \times \Theta \left (\alpha < \sqrt{\frac{k_{a}\cdot k_{j}}{k_{j}\cdot q}\frac{k_{b}\cdot q}{k_{a}\cdot k_{b}}} q^{(ab)}_{\bot} < \beta \right), \label{eqn:18}
\end{align}
where $j$ labels a final-state particle. Written this way, we see that $\Omega_{a j} \neq \Omega_{b j}$ for all $j$. This means that almost all observables at hadron-hadron colliders that have any sensitivity to soft gluon emission will violate coherence to some degree. As pointed out in \cite{Becher:2021zkk}, this includes Drell-Yan and $gg \rightarrow H$ hard-processes, since their colour can become sufficiently involved after emitting two or more gluons into the final state (i.e. coherence violation will first appear in $\frac{\td \sigma_2}{\td x_a \td x_b}$).

\section{The logarithmic order of coherence violation} 

For the majority of pure QCD observables, coherence violation will emerge for the first time at $\mathcal{O}(\As^4)$ in the fixed order expansion (relative to the order of the Born process). That's because one needs at least one soft gluon, one collinear emission and two Coulomb exchanges.

The logarithmic order at which coherence violation will occur is process dependent. We consider a general measurement function which produces logarithms $\ln v^{-1} \equiv L$: 
\begin{align} u(\{k\}) = \sum_{j} F_{j}(\{k\}) \Theta (v - V_{j}(\{k\})).\end{align}
Observables for which $F_{j}= 1$ are known as event-shape observables \cite{Banfi:2004nk,Resum_large_logs_ee,Korchemsky:1999kt} and observables for which $F_{j} \neq 1$ are weighted cross-sections \cite{Sveshnikov:1995vi,Tkachov:1995kk,Korchemsky:1999kt,Hofman:2008ar,Chen:2020vvp}. We will give specific examples of the functions $F_j$ and $V_j$ below. To get the leading coherence-violating logarithm we must take the $z \rightarrow 1$ limit of Eq.~\eqref{eq:start}. As anticipated, the first potentially non-vanishing term occurs at $\mathcal{O}(\As^4)$ relative to the Born result:
\begin{align}
&\frac{\td \sigma_1}{\td x_a \td x_b\td \mathcal{B}} \approx \nonumber \\ &\sum_{\substack{i = a,b \\ j \neq a,b}} A^{(1234)}_{ij} \int^{Q}_{\mu_{\TT{F}}} \frac{\td k_{4 \bot}}{k_{4 \bot}} \left[ \int^{Q}_{k_{4 \bot}} \frac{\td k^{(ab)}_{3 \bot}}{k^{(ab)}_{3 \bot}} \int \frac{\td y_{3}\td \phi_{3}}{2\pi } w_{ij}\right] \left( \int^{Q}_{k^{(ab)}_{3 \bot}} \frac{\td k_{2 \bot}}{k_{2 \bot}}\int^{1}_{k_{2 \bot}/Q}\frac{\td \theta_2}{\theta_2} \right)    \nonumber \\
& \qqquad \times \int^{Q}_{k_{2 \bot}} \frac{\td k_{1 \bot}}{k_{1 \bot}} \; u_{1}(k_{2})(1 - u_{2}(k_{3},k_{2})) + (1243) + (2134) + (2143) + (2314). \label{eq:banfi}
\end{align}
The collinear parton is parton 2, parton 3 is a soft wide-angle gluon and partons 1 and 4 are Coulomb exchanges, $w_{ij} = (k^{(ab)}_{3 \bot}/k^{(ij)}_{3 \bot})^{2}$ and
$$
A^{(1234)}_{ij} = \left( \frac{\As}{\pi}\right)^4 C^{(1234)}_{ij} \,(i\pi)^{2} \, f_{A}(x_{a}, \mu_{\TT{F}})f_{B}(x_{b}, Q),
$$
where
$$
C^{(1234)}_{ij} = \Tr \left(\left[\v{T}^{2}_{s}, \v{T}_{i}\cdot \v{T}_{j} \right] (\v{T}_{a} [\v{T}^{2}_{s} ,\v{H}] \v{T}^{\dagger}_{a} - \v{T}^{2}_{a} [\v{T}^{2}_{s} ,\v{H}] ) \right).
$$
The four additional terms indicated on the second line correspond to different ways of ordering the parton transverse momenta: the main term corresponds to $(1234)$ and, for example, $(2134)$ corresponds to the collinear parton having the largest $k_\bot$ etc. The colour factors for the other orderings are
$$
C^{(1243)}_{ij} = C^{(1234)}_{ij},
$$
and
\begin{align}
    C^{(2134)}_{ij} &= C^{(2143)}_{ij} = C^{(2314)}_{ij} = -\Tr \left(\left[ \v{T}^{2}_{s} , \left[\v{T}^{2}_{s}, \v{T}_{i}\cdot \v{T}_{j} \right]\right] (\v{T}_{a}\v{H}\v{T}^{\dagger}_{a}-\v{T}^{2}_{a} \v{H}) \right). \nonumber 
\end{align}
As parton 3 is a wide-angle gluon its angular integrals generate observable-dependent, finite but not-logarithmically enhanced terms when restricted by the two parton measurement function, $1-u_2(k_{3},k_{2})$. Though these terms may not factorise from Eq.~\eqref{eq:banfi}, they can be ignored in the subsequent discussion as they do not effect the logarithmic power counting. In Eq.~\eqref{eq:banfi} we have used that in the soft-collinear limit $k_{2 \bot}/E_{a} \approx (1-z)\theta_2$ where $z$ is the momentum fraction used in Eq.~\eqref{eq:start}. $\mu_{\TT{F}}$ is the factorisation scale, below which proton evolution is completely DGLAP. It is chosen to have the largest value such that $u_{2}(k_{3},k_{2}) \approx 1$ given $k_{3 \, \bot}<\mu_{\TT{F}}$, e.g. in gaps-between-jets $\mu_{\TT{F}}=Q_0$. This factorisation scale choice naturally extends to all orders where we require that $\mu_{\TT{F}}$ has the largest value such that $u(\dots, k_{i} , k_{s}, k_{j} , \dots) \approx u(\dots, k_{i} , k_{j}, \dots)$ when  $k^{(ab)}_{s \, \bot}<\mu_{\TT{F}}$ for all wide-angle soft momenta $k_{s}$.

We are interested in determining the logarithmic behaviour of Eq.~\eqref{eq:banfi}. This is determined by $u_{1}(k_{2})(1 - u_{2}(k_{3},k_{2}))$. There are three scenarios that we must study when evaluating Eq.~\eqref{eq:banfi}. Firstly we can consider when $\max (k_{2 \bot}) \gg \mu_{\TT{F}}$, where $\max (k_{2 \bot})$ is the smallest value of $k_{2 \bot}$ such that for $k_{2 \bot} > \max (k_{2 \bot})$ both $u_{1}(k_{2}) \approx 0$ and $u_{2}(k_{3},k_{2}) \approx 0$. In this situation, each of the 5 nested integrals generates a logarithm (the infra-red safety of the functions $F_{i}$ means they do not alter the logarithmic counting in this limit). As a result, $\frac{\td \sigma_1}{\td x_a \td x_b} \sim \As^4 L^5$. Secondly we have the case $\max (k_{2 \bot}) \approx w \, \mu_{\TT{F}}$, where $w \gtrsim 1$. This means that the observable restricts the phase-space in the collinear region such that upper limit on $k_{2 \bot}$ in Eq.~\eqref{eq:banfi} can be exchanged with $w \, \mu_{\TT{F}}$. Consequently integrals over $\td k_{4 \bot} \td k^{(ab)}_{3 \bot}\td k_{2 \bot}$ generate a term proportional to $(\ln w)^{3} \sim \mathcal{O}(1)$. Logarithms are still produced since either the $\td k_{1 \bot}$ integral generates a single logarithm or both the $\td k_{1 \bot}$ and the $\td \theta_2$ integrals generate logarithms. As a result, $\frac{\td \sigma_1}{\td x_a \td x_b} \sim \As^4 L ~ (\ln w)^{3}$ or $\frac{\td \sigma_1}{\td x_a \td x_b} \sim \As^4 L^2 ~ (\ln w)^{3}$ respectively. We cannot rule out that a complete calculation finds $\ln w = 0$. However, any such cancellation would be observable dependent and determined by final-state kinematics. Examples of both $\As^4 L (\ln w)^{3}$ and $\As^4 L^2 (\ln w)^{3}$ observables are given in the following paragraph. Note that in this case, the $(2jkl)$ terms (where $j,k,l \in \{1,3,4\}$) are sub-leading and thus the logarithms have smaller numerical prefactors, since fewer topologies contribute. Finally, there is the case $\mu_{\TT{F}} = \max (k_{2 \bot})$. This can only occur if $u_{1}(k_{2})(1 - u_{2}(k_{3},k_{2})) = 0$ for all $k^{(ab)}_{3 \bot} \sim k_{2 \bot}$, which means the observable is completely insensitive to wide-angle radiation and so the hadron evolution can be completely described using DGLAP evolution without soft resummation. In other words, the observable is trivially without coherence violating logarithms. Observables of this form include the modified massdrop tagger \cite{Dasgupta:2013ihk} and the collinear limit of $N$-point energy correlators \cite{Chen:2020vvp} applied within massless jets.

To illustrate matters, we will review coherence violation in continuously-global observables. For these, the measurement function can be written $u = F(\{p\}) \Theta (v - V(\{p\}))$, where $V(\{p\})$ and $F(\{p\})$ are defined to have the following properties \cite{Banfi:2004yd}:  
\begin{itemize}
	\item For a single, soft emission, $k$, that is collinear to hard parton $i$, $$V(\{p\}) = d_{i} \left(\frac{k^{(i n)}_{\bot}}{Q}\right)^{h}e^{-l_{i} y_{k}}g_{i}(\phi_{k}),$$ where $d_{i},h,l_{i}$ are constants, and $g_{i}(\phi_{k})$ can be any function of the azimuth for which the integral $\int \td\ln\phi_{k} \;  g_{i}(\phi_{k})$ exists. $k^{(i n)}_{\bot}$ is the transverse momentum relative to parton $i$ and any other arbitrary direction given by the unit vector $\vec{n}$. In the limit that $k$ is both soft and collinear to $i$, the choice of $\vec{n}$ is sub-leading. To be global, all of the $d_i \neq 0$.
	\item For a single, soft emission, $k$, that is not collinear to any hard parton, $$V(\{p\}) \sim \left(k^{(ab)}_{\bot}\right)^h,$$ where $h$ has the same value as in the collinear case above. This ensures the observable's scaling in transverse momentum is continuous across all logarithmically enhanced regions of phase-space.
	\item $F \sim 1 + A (k^{(ab)}_{\bot}/Q)^{h'}$ where $h'>0$ and $A$ are constant over the entire phase space of a soft parton with momentum $k$.
\end{itemize}
The effects of coherence violation on continuously-global observables were first evaluated for event shape observables in \cite{Banfi:2010xy}. The continuous scaling in transverse momentum of these observables allows us to set   $\mu_{\TT{F}} \approx Q e^{-\frac{L}{h}}$ and  $u_{2}(k_{3},k_{2}) \rightarrow 0$. If the collinear parton is soft and collinear to parton $a$ and $V(k_{2}) \sim (k_{2 \bot}/Q)^h \theta^{l_a}_2$, we can replace $$u(k_{2}) \rightarrow \Theta(k_{2 \bot} \theta^{\frac{l_{a}}{h}}_2 \lesssim Q e^{-\frac{L}{h}}).$$ Thus Eq.~\eqref{eq:banfi} gives $\frac{\td \sigma_1}{\td x_a \td x_b} \sim \As^4 L (\ln w)^3$ for $l_a<0$, and $ \frac{\td \sigma_1}{\td x_a \td x_b}\sim \As^4 L^2 (\ln w)^3$ for $l_a=0$. When $l_a>0$ every term contributes and $\frac{\td \sigma_1}{\td x_a \td x_b} \sim \As^4 L^5$. In \cite{Banfi:2010xy} it was identified that $l_a \leq 0$ is the case for `standard' rIRC observables (such as transverse-thrust, for which $l_{a,b}=0$), whereas $l_a > 0$ typically occurs in `exponentially-suppressed' rIRC observables.

For example, we can compute the cumulative scalar and vector sum $p_{T}$ distributions for $ p p \rightarrow W^{\pm},Z,H$ processes.  The measurement function for the vector sum cumulative distribution is determined by momentum conservation \cite{Ellis:1991qj}:
\begin{align}
    u(k_{1},\dots, k_{n}) = \Theta\left(\vec{p}_{\bot}^{\; 2} - 
 \left(\sum^{n}_{i=1}\vec{k}_{i \, \bot}\right)^{2}  \right)
\end{align}
where $\vec{p}_{\bot}^{\; 2}$ is the upper limit on the squared $p_{T}$ of the colour-neutral boson. In contrast, the scalar sum distribution is given by the measurement function:
\begin{align}
    u(k_{1},\dots, k_{n}) = \Theta\left( p_{\bot}^{\; 2} - \sum^{n}_{i=1} \left(  \vec{k}_{i \, \bot}\right)^{2}  \right)
\end{align}
where the parameter $p_{\bot}^{\; 2}$ does not directly relate to the kinematics of the final state boson. To leading order, the measurement function for the scalar sum distribution has the form studied in the previous paragraph with $l_a=0$ and $L = \ln M^{2}_{W^{\pm},Z,H}/p_{\bot}^{\; 2}$. As the Born final state is a colour singlet, a second collinear emission must be inserted into the strongly ordered cascade in order for the $\v{T}_{s}^{2}$ colour commutators in $C_{ij}$ to not vanish. The leading CVL is given by the $(2jkl)$ terms in Eq.~\eqref{eq:banfi}, where $j,k,l \in \{1,3,4\}$, with the second collinear emission inserted either immediately before or after parton 2\footnote{For this CVL $$C_{ij}= -\sum_{a'}\Tr \left(\left[ \v{T}^{2}_{s} , \left[\v{T}^{2}_{s}, \v{T}_{i}\cdot \v{T}_{j} \right]\right] (\v{T}_{a'}\v{h}\v{T}^{\dagger}_{a'}-\v{T}^{2}_{a'} \v{h}) \right)$$ where $\v{h}= \v{T}_{a} \v{H}\v{T}^{\dagger}_{a}  - \v{T}^{2}_{a} \v{H}$ and where the sum over $a'$ includes parton $a$ and the collinear radiation emitted from parton $a$.}. Consequently, the lowest order coherence violating logarithm in the scalar sum distribution will be of the form $\frac{\td \sigma_2}{\td x_a \td x_b} \sim \As^5 L^{2} (\ln w)^4$. For the scalar sum distribution, we see no reason to anticipate a cancellation which could lead to $\ln w = 0$. The computation of the vector sum is much the same and we once again find a possible CVL of the form $\frac{\td \sigma_2}{\td x_a \td x_b} \sim \As^5 L^{2} (\ln w)^4$. However, in this instance the seminal CSS proof of Drell-Yan factorisation \cite{Collins:1984kg,Collins:1987pm} indicates that no CVL will be present. Translated into the vernacular of this paper, CSS demonstrated that, since the measurement applies no constraint on QCD radiation beyond that of momentum conservation, the complete momentum conserving delta function forces $\ln w = 0$. 

Finally, we consider not continuously-global observables. For these $h$, is not constant over the entire phase space and $F$ need not go like $\sim 1 + A(k_\bot^{ab}/Q)^{h'}$. This means that the phase-space of an emission can be divided into at least two regions, $c$ and $s$, between which the scaling of the observable differs. For example, the gaps-between-jets observable is such that, in the jet regions, $h =0$ but in the gap region $h \approx 1$. Infra-red safety dictates that each region, $c$ or $s$, has an inclusivity scale, $\mu_{c}$ or $\mu_{s}$, such that the observable is insensitive to radiation emitted into that region with $k_{t} \lesssim \mu_{c,s}$. Let $c$ contain the region collinear to parton $a$ and $s$ be the  complimentary region. Therefore in Eq.~\eqref{eq:banfi} $u(k_{2}) \approx 1$ for $k_{2 \bot}<\mu_{c}$ whilst $u(k_{2},k_{3}) \approx u(k_{2})$ for $k^{(ab)}_{3 \bot} < \min(\mu_{c},\mu_{s})$, since a soft emission can be emitted into either of the regions $c$ and $s$. Hence $\mu_{\TT{F}} = \min(\mu_{c},\mu_{s})$, e.g. for gaps-between-jets $\mu_{c}=Q$, $\mu_{s}=Q_{0}$ and therefore $\mu_{\TT{F}}=Q_{0}$ as per Eq.~\eqref{eq:master}. The inclusivity scales are functions of the parameters defining the observable. As the observable is not continuously-global, $\mu_{c}$ and $\mu_{s}$ do not scale proportionally to each other under variation of those parameters. Again consider gaps-between-jets where the parameters $Q_{0}$ and $Y$ are used to define the observable: $\mu_{s}$ is linear under the variation of $Q_{0}$ whilst $\mu_{c}$ is a constant\footnote{Both $\mu_{c}$ and $\mu_{s}$ are constant under the variation of $Y$ which only determines the angular extent of the regions $c$ and $s$}. Necessarily, as a consequence of unitarity in the collinear region around parton $a$, $\max (k_{2 \bot}) \approx \mu_{c}$. Therefore, in the resummation limit of a not continuously-global observable, either $\max (k_{2 \bot}) \gg \mu_{\TT{F}} = \mu_{s}$ or $\max (k_{2 \bot}) \approx \mu_{\TT{F}} = \mu_{c} \ll \mu_{s}$. When $\max (k_{2 \bot}) \gg \mu_{\TT{F}}$, $\frac{\td \sigma_1}{\td x_a \td x_b}\sim \As^4 L^5$ and so the observable suffers coherence violating logarithms. As before, when $\max (k_{2 \bot}) = \mu_{\TT{F}}$ the observable is trivially insensitive to soft radiation. 

\section{Conclusions}

Our analysis shows that in hadron colliders almost all observables with sensitivity to wide-angle soft radiation dressing the initial state hadrons will suffer coherence violation. In the previous section, we computed the logarithmic order of this violation at fixed order in $\As$. We see no arguments for why our analysis cannot be extrapolated to $n$th order in perturbation theory\footnote{Higher order calculations have been performed in \cite{Keates:2009dn,Becher:2021zkk}, see also \cite{Forshaw:2019ver}.}. Provided the observable under consideration has leading logarithms of the form $\As^{n} L^{2n}$ and the Born hard process is real, we expect coherence violating logarithms of the form:
\begin{itemize}
	\item $\As^{n} L^{2n-6}$ or $\As^{n} L^{2n-7}$  for $n \geq 4$ in standard continuously-global observables \cite{Banfi:2004yd,Banfi:2004nk} where the Born hard process has a coloured final state. These contribute at the same accuracy as those in an N$^{3}$LL or N$^{4}$LL exponentiation. Consequently, we do not find conflict with the N$^{2}$LL resummations of global observables in $pp$ collisions \cite{Bozzi:2005wk,Berger:2010xi,Stewart:2010qs,Stewart:2010tn,Stewart:2010pd,Banfi:2012jm,Stewart:2013faa,Catani:2014qha,Becher:2015gsa} including those which were derived in SCET without the inclusion of Glauber modes \cite{Stewart:2010qs,Stewart:2010tn,Stewart:2010pd,Stewart:2013faa,Becher:2015gsa}. When the Born hard process has a colour singlet final state, standard continuously-global observables generate either $\As^{n} L^{2n-8}$ or $\As^{n} L^{2n-10}$ terms for $n \geq 5$. 
	\item $\As^{n} L^{2n-3}$ for $n \geq 4$ in `forward suppressed' continuously-global observables (with  $l_a>0$ as defined in the previous section). Though, as these logarithms first emerge at $\mathcal{O}(\As^4)$ they will contribute at the same accuracy as those in a LL exponentiation. For colour singlet Born final states CVL are only present for $n \geq 5$.
	\item $\As^{n} L^{2n-3}$ for $n \geq 4$ for not continuously-global observables. When a not continuously-global observable has leading logarithms of the form $\As^{n} L^{n}$, coherence violating logarithms will become superleading. This case has recently been resummed for the first time \cite{Becher:2021zkk}. For colour singlet Born final states CVL are only present for $n \geq 5$.
\end{itemize}
We note that CVLs less enhanced than $\As^{n} L^{2n-4}$ are accompanied by an additional $\mathcal{O}(1)$ observable-dependent kinematic factor whose calculation is beyond our control ($\ln w$ in the previous section). We cannot rule out that this factor is zero and it is zero for the $W, Z, H$ boson $p_T$ distributions, where momentum conservation engineers a cancellation \cite{Collins:1987pm,Collins:1984kg}.

A remark on the role of electroweak hard processes. In \cite{Catani:2011st,factorisationBreaking} it was noted that two Coulomb exchanges are not needed to ensure real coherence-violating terms emerge in resummations dressing electroweak hard-processes (for instance the hard process is the sum of $s$ and $t$ channel amplitudes for $qq'\rightarrow qq'$ hard processes mediated by W or Z bosons). This is because the hard process itself can supply a complex phase, which allows terms with a single Coulomb exchange to contain a real piece that can contribute to the cross-section. Thus for such hard-processes there is a possibility for $\mathcal{O}(\As^3)$ coherence-violating logs to emerge as well as the $\mathcal{O}(\As^4)$ upwards which we have studied. By repeating the analysis of the previous section, we see that in this case coherence violation could contribute logarithms of the form $\As^{n} L^{2n-2}$ for $n \geq 3$. For continuously global observables this would give CVLs of the form $\As^{n} L^{2n-5}$ or $\As^{n} L^{2n-6}$ for $n \geq 3$. As before, when the Born hard process has a colour singlet final state, an additional factor of $\As$ is required, leading to CVLs of the form $\As^{n} L^{2n-7}$ or $\As^{n} L^{2n-9}$ for $n \geq 4$.

\section{Acknowledgements}

JRF thanks the Mainz Institute for Theoretical Physics (MITP) (Project ID 39083149) for its hospitality and support. JH thanks the UK Science and Technology Facilities Council for the award of a postgraduate studentship. This work is supported in part by the GLUODYNAMICS project funded by the``P2IO LabEx (ANR-10-LABX-0038)'' in the framework ``Investissements d’Avenir'' (ANR-11-IDEX-0003-01) managed by the Agence Nationale de la Recherche (ANR), France. We would like to thank Simon Pl\"atzer, Mike Seymour, Aditya Pathak and Gavin Salam for insightful discussions. We would also like to thank Jack Helliwell, Matthew De Angelis and the Manchester QCD theory group for broad and helpful discussions.

\bibliographystyle{JHEP}
\bibliography{SLL-letter}

\end{document}